# SPARSE BAYESIAN STEP-FILTERING FOR HIGH-THROUGHPUT ANALYSIS OF MOLECULAR MACHINE DYNAMICS

*Max A. Little[*a], Nick S. Jones[*a]*

[a]Oxford Centre for Integrative Systems Biology, Department of Physics, Oxford University, UK

**ABSTRACT**

Nature has evolved many molecular machines such as kinesin, myosin, and the rotary flagellar motor powered by an ion current from the mitochondria. Direct observation of the step-like motion of these machines with time series from novel experimental assays has recently become possible. These time series are corrupted by molecular and experimental noise that requires removal, but classical signal processing is of limited use for recovering such step-like dynamics. This paper reports simple, novel Bayesian filters that are robust to step-like dynamics in noise, and introduce an $L_1$-regularized, global filter whose sparse solution can be rapidly obtained by standard convex optimization methods. We show these techniques outperforming classical filters on simulated time series in terms of their ability to accurately recover the underlying step dynamics. To show the techniques in action, we extract step-like speed transitions from *Rhodobacter sphaeroides* flagellar motor time series. Code implementing these algorithms available from http://www.eng.ox.ac.uk/samp/members/max/software/.

*Index Terms*—molecular machines, digital filter, Bayes theorem, L1-regularization, convex optimization

## 1. INTRODUCTION

Nanotechnology promises extraordinary control over matter at the molecular scale, with widespread applications from materials to medicine. However, constructing useful machines at this scale is enormously challenging because of molecular noise and the complexity of engineering precise conformational dynamics of interacting molecular components. Nature has evolved many robust molecular machines such as pumps, tugs, copiers and motors, and understanding the function of these machines offers the possibility of the biomimetic design of interacting, artificial molecular devices. Biophysical theory proposes that these motors, that convert electrochemical to linear or rotary kinetic energy, do so in a series of rapid, nano-scale, step-like motions because this maximizes the use of the available free energy in chemical bonds [1]. This step-like motion has recently been observed, in vivo, using advanced experimental assays exploiting techniques such as Förster resonance energy transfer, optical traps or atomic force microscopy [2].

These assays produce large volumes of time series data with sampling rates that often exceed 100kHz, but the step-like motions of the molecular components is inevitably obscured by Brownian and experimental noise [1, 3]. This noise must be removed to extract the underlying molecular conformational dynamics. This noise removal is a challenging signal processing problem, because the signal-to-noise ratio is low and the signal itself is step- and impulse-like, i.e. it is *highly discontinuous*. Thus, the time scale for changes in the noise and the signal are similar, so that the support in the Fourier domain of the signal and the noise overlap considerably. Separation in the Fourier domain using classical linear filtering is, therefore, largely intractable. Special techniques are therefore required that can cope with both high noise levels and discontinuous signals.

Although special step-smoothing filters do exist [4], they have various shortcomings. Nonlinear adaptations to the classical running mean filter [5] – given the fundamental problem of frequency inseparability of steps from noise – lack statistical robustness. Others are based on *greedy* (locally optimal), successive subdivision search for best step locations, but they are computationally intensive and the solutions they find may be suboptimal [6, 7]. There has also been much recent work on hidden Markov modeling to solve similar filtering problems [8], but the computational complexity of statistical inference with these techniques is substantial, and generally prohibitive for high-throughput experimental situations.

There is therefore the need for robust step-smoothing filters that produce optimal solutions with minimal computational effort and good statistical power.

## 2. METHODS

### 2.1. Step-like synthetic time series

Typical noisy step-like time series from experimental assays of molecular machines is depicted in Fig. 1(a). The prototypical time series is a series of piecewise-constant segments with superimposed, additive, i.i.d. Gaussian noise. The steps between the constant segments typically occur at time intervals that are exponentially distributed, and the steps may be upwards or downwards. This is very similar to a Poisson process, except that the event count can go down as well as up. The observed discrete-time signal is defined here as $x_n = \mu_n + \varepsilon_n$, where $\mu_n$ is a piecewise-constant step-signal with steps of the same height (here we assume unit height without loss of generality), and $\varepsilon_n$ is i.i.d. Gaussian noise of variance $\sigma^2$.

[*] Funded by the BBSRC and the EPSRC, UK.

## 2.2. Step-filtering algorithms

As a simple reference algorithm that is very commonly used in this application, we investigate the classical *running median filter* [9], which replaces the middle sample of a moving window that runs through the time series with the *median* of the samples in that window. The only parameter is the window length *W*. This filter is a potentially useful candidate for step-filtering because it is computationally and conceptually very simple, but, unlike the *running mean filter*, it leaves edge- and impulse-like *root signals* unchanged. The filter will leave only root signals if the signal is run through the filter sufficient number of times. Another justification for the operation of the filter is that it is the maximum-likelihood estimate of the *location*, *m* of the distribution of samples in the window, if they are Laplace-distributed. The negative likelihood function of the window samples is therefore:

$$-\ln P(x_w|m) = -W \ln A + a \sum_{i \in w} |x_i - m| \qquad (1)$$

where *w* is the size *W* index set of samples in each window, *A* is an unimportant normalization factor, and *a* is the spread of the Laplace distribution. Minimizing this function with respect to *m* is equivalent to minimizing $E = \sum_{i \in w} |x_i - m|$, which is solved when *m* is the median of the samples [10].

The running median filter has some shortcomings that make it less than ideal for step-filtering. In particular, the root signals include monotonic sections (*up/down ramps*) that cannot feature in $\mu_n$ by definition. As an improvement, if we assume we know the step positions, we can place a Bayesian Laplace *mixture prior* over the estimate *m* for each window:

$$P(m) = \sum_{j=1}^{S} B \exp(-b|m - s_j|) \qquad (2)$$

with spread *b* around each of the *S* step positions $s_j$, and (unimportant) normalization factor *B*. Then, combining this with the likelihood (1), the negative log posterior is:

$$-\ln P(m|x_w) = a \sum_{i \in w} |x_i - m| - \ln \sum_{j=1}^{S} \exp(-b|m - s_j|) + K \qquad (3)$$

where *K* is an unimportant constant that can be ignored when finding the *maximum a-posteriori* (MAP) solution *m* for each window (because it does not depend on *m*). Unfortunately, since this negative log posterior is not a *convex function*, standard minimization algorithms are not guaranteed to find the global minimum solution. However, since there is only one parameter *m*, a first-pass brute-force search for a rough estimate of *m* can be combined with a second-pass golden-section search refinement [11]. This is guaranteed to converge on the optimum if it is within the range of the first-pass search.

The above filter requires knowledge of the step positions; in certain experimental situations these are known, for example, high-throughput DNA sequencing by nanopore blockade current. In other circumstances, step positions will be unknown. Furthermore, the non-convexity of the negative log posterior makes finding the optimal *m* difficult. Also, this windowed algorithm may fail to smooth out steps longer than the window size or erroneously smooth away multiple steps within a single window.

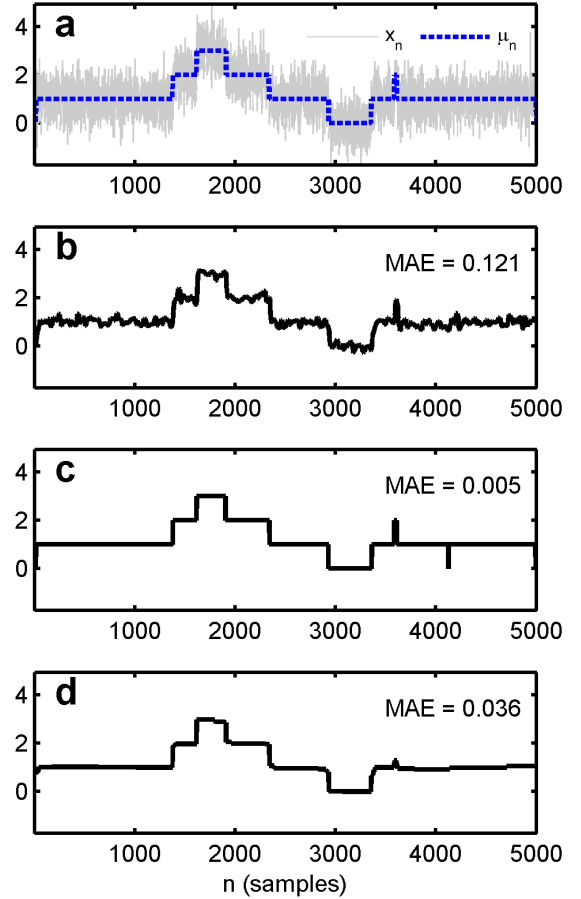

**Fig. 1. (a)** Typical time series from experimental assays of molecular machines, showing Poisson step-like dynamics (unit step positions) obscured by additive Gaussian noise of variance $\sigma^2 = 0.36$. **(b)** Median filtering with window size *W* = 20, **(c)** Bayesian Laplace mixture prior median filtering with window size *W* = 20, *a* = 0.1 and *b* = 10, assuming unit step positions $s_j$, and **(d)** Bayesian $L_1$-regularized fused-LASSO global filtering with regularization parameter $\lambda = 10$. MAE in each panel is the Mean Absolute Error (see text) which quantifies the accuracy of each step-filtering algorithm, smaller is better.

A different approach is offered by *global filtering* that finds an optimal solution for the whole time series at once, rather than considering a sliding window of samples. It is reasonable to assume that the additive error around each step is Gaussian, and thus we assume that the likelihood is Gaussian. Furthermore, since the most prominent property of step-filtering is that many of the underlying, adjacent step samples $\mu_n$ take on the same values, the $L_1$-regularized fused-LASSO [12] offers a plausible Bayesian method. This prescription allows us to write down the following negative log posterior:

$$-\ln P(m|x) = \sum_{n=1}^{N-1}(x_n - m_n)^2 + \lambda \sum_{n=1}^{N-1}|m_{n+1} - m_n| + K \quad (4)$$

where $\lambda$ is a *regularization* parameter, $N$ is the number of samples in the time series, and $K$ is another unimportant constant. For $\lambda = 0$, the optimal solution $m$ must be identical to $x$, and as $\lambda \to \infty$, all adjacent $m$'s take on the same value, so $m_n \to k$, some arbitrary constant. For all finite values of $\lambda$, the solution is a piecewise-constant curve with a finite number of steps, that is simultaneously the least-squares fit to the data. This piecewise-constant property emerges because most of the differences between adjacent $m$'s will be zero: this is the recently described *sparsity enhancing* property of the Laplacian prior [12] of exemplary value in compressive sensing applications. A convenient aspect of (4) is that this is a convex *quadratic programming* problem guaranteeing that a globally optimal solution can be rapidly found to arbitrary precision using standard algorithms [10]. Note that (4) is similar to the piecewise-linear smoothing algorithm proposed by Kim *et al.*[13].

**Table 1. Accuracy of the three step-filtering algorithms in terms of Mean Absolute Error (MAE), in recovering synthetic underlying step dynamics $\mu_n$, for unit step height, for a range noise variances. Algorithm parameters are as detailed in the caption to Fig. 1.**

| Gaussian i.i.d. noise variance $\sigma^2$ (95% interval in brackets) | Median filter MAE | Bayesian Laplace mixture prior median filter MAE | Bayesian $L_1$-regularized fused-LASSO global filter MAE |
|---|---|---|---|
| 0.01 (0.2) | 0.035 | 0.006 | 0.014 |
| 0.09 (0.6) | 0.072 | 0.006 | 0.028 |
| 0.25 (1.0) | 0.103 | 0.006 | 0.029 |
| 0.49 (1.4) | 0.127 | 0.016 | 0.046 |
| 0.81 (1.8) | 0.174 | 0.047 | 0.066 |
| 1.21 (2.2) | 0.206 | 0.086 | 0.074 |

### 2.3. Quantifying algorithm performance

The accuracy of each of the three algorithms – 1. median filtering, 2. Bayesian Laplace mixture prior median filtering, and 3. Bayesian $L_1$-regularized fused-LASSO global filtering – is assessed by computing the Mean Absolute Error $MAE = \sum_{n=1}^{N}|\mu_n - m_n|/N$ of the estimate of underlying Poisson step-like dynamics.

These three algorithms have several (hyper-)parameters that must be chosen, for example, the window size $W$ and the regularization parameter $\lambda$. Here, to make a fair comparison of the performance of these algorithms, we choose these parameters such that the MAE in recovering the known step-like dynamics is minimized. In practice, these parameters would be chosen, using, for example, an appropriate cross-validation scheme.

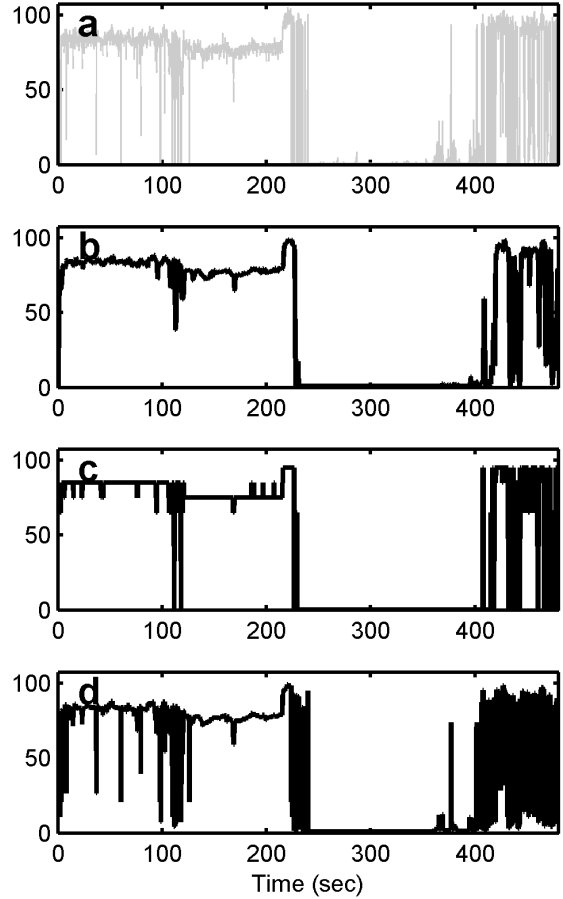

**Fig. 2. (a)** Typical time series of *R. sphaeroides* bacterial motor rotation speed against time. **(b)** Median filtering with window size $W = 10$. **(c)** Bayesian Laplace mixture prior median filtering with window size $W = 10$, $a = 0.1$ and $b = 10$, assuming step positions $s = \{0, 65, 75, 85, 95\}$, **(d)** Bayesian $L_1$-regularized fused-LASSO global filtering with regularization parameter $\lambda = 10$. Vertical axis is rotation speed in Hz.

### 3. EXPERIMENTAL DATA

The data for this study comes from WS8N wild-type *Rhodobacter sphaeroides* cells. The flagella (tails) are removed, and 0.83 micron beads are attached to the flagellar hook. The beads are then laser illuminated, and the speed of rotation of the flagellar motor against time is recorded. Fig. 2(a) depicts a typical time series from this experimental assay, which is described fully in Pilizota *et al.* [14].

### 4. RESULTS AND DISCUSSION

Example synthetic time series and the algorithm estimates are shown in Fig. 1(b), (c) and (d). As can be seen, the Bayesian Laplace mixture prior median filter is the most accurate up to a noise variance of around 0.8. However, this filter requires strong assumptions about the step positions, and usually this information is not known in advance. The Bayesian $L_1$-regularized fused-LASSO global filter is the next most accurate up to noise variance of 0.8, beyond which it is the most accurate. Both of these two novel algorithms produce usefully smooth results with sharp edges, whereas the median filter produces noisy results and the edges lack definition. However, both the median and Bayesian median filters retain the fast, impulse-like change seen at ~3700 samples in Fig. 1.

The performance of the algorithms on synthetic stepping time series with unit step height across a range of noise variances is shown in Table 1. The median filter has the worst overall performance, as the error can reach as much as 20% of the step height. By contrast, the two novel Bayesian filters can readily achieve errors of less than 10% of the step height.

Fig. 2. shows a typical experimental time series and the results of applying the three step-smoothing filters. As can be seen, the time series has fairly long periods of constant rotation (obscured by noise), but it also has periods where the speed changes rapidly from stopped, to upwards of 100Hz. Thus, noise removal is extremely challenging for classical running filters such as the median filter, which cannot generally detect this situation where several large changes are occurring within a window. The Bayesian adaptation to the median filter performs better: it not only produces the smoothest estimates of the constant rotation periods, but responds better to the rapid speed change events than the median filter. The $L_1$-regularized fused-LASSO global filter is capable of responding to the rapid speed changes, but is not as effective as the Bayesian median filter at smoothing the constant rotation periods.

## 5. SUMMARY AND CONCLUSIONS

We studied noise removal for time series recorded from experimental assays of the step-like dynamics of molecular machines. We introduced two novel Bayesian filters, both of which outperformed the classical running median filter in accurate recovery of the underlying stepping behaviour in synthetic time series when corrupted by noise of increasing variance. We also demonstrated that these two new filters are capable of recovering rapid stepping combined with piecewise-smooth segments obscured by noise, but the classical median filter fails in this situation. Given the poor accuracy of the median filter, it is only useful where the noise has low variance relative to the step height. Overall then, we might prefer the Bayesian $L_1$-regularized fused-LASSO filter, because it produces results of similar accuracy to the Bayesian Laplace mixture prior median filter, but has only one hyperparameter.

As demonstrated here, analysing time series from molecular machines opens up interesting new challenges in signal processing. Furthermore, the volume of experimental data being generated is growing at an exponential rate. Classical statistical signal processing tools are simple and computationally cheap, and thus well-suited to high-throughput analysis, but the mathematical assumptions of traditional statistical signal processing are fundamentally inappropriate for step- and impulse-like behaviour corrupted by high noise levels typical of time series from these experimental assays. Our aim in this paper has been to design novel methods that combine the best of both worlds: computational robustness and simplicity with accuracy for the problem at hand.

## 6. ACKNOWLEDGEMENTS

The authors would like to thank Mostyn Brown of the Department of Biochemistry, Oxford University, UK for providing the *R. sphaeroides* rotation speed data and for fruitful discussions and advice.